\documentclass[twocolumn]{revtex4}
\usepackage{graphicx}
\begin{document}
\title{Atoms, dipole waves, and strongly focused light beams}
\author{S.J. van Enk}
\affiliation{Bell Labs, Lucent Technologies\\
600-700 Mountain Ave,
Murray Hill, NJ 07974}
\date{\today}

\begin{abstract}We describe the resonant interaction of an atom with a
strongly 
focused light beam by expanding the field in multipole waves.
For a classical field, or when the field is described by a coherent state,
we find that both intensity pattern and photon statistics
of the scattered light are fully determined by a small 
set of parameters. One crucial parameter is
the overlap of the field with the appropriate dipole wave corresponding to
the relevant dipole transition in the atom.
We calculate this overlap for a particular set of strongly 
focused longitudinally polarized
light beams, whose spot size is only $0.1\lambda^2$,
as discussed in S.~Quabis {\em et al.}, 
Appl. Phys. B {\bf 72}, 109 (2001). 
\end{abstract}

\maketitle
\section{Introduction}
For many applications, both in classical and quantum optics,
it is a good idea to focus light to a small
spot size. 
For example, in recent
experiments on single atoms trapped in tiny dipole traps \cite{grangier}
the focusing system plays a major role. 
A natural question in that context is, how strong can the 
interaction between a single atom 
in free space and a light wave be? 
One expects the strength of the interaction to increase
with decreasing focal spot size. However, there must be an upper limit 
to the interaction strength for at least two reasons: first, 
there is a limit to how much one can focus 
light of a given wavelength $\lambda$; second, the cross section of 
a two-level atom is $\sigma=3\lambda^2/(2\pi)$, which 
seems to indicate that as long as all
light is focused to within $\sigma$ the interaction is optimal.
The latter argument, however, does not paint the complete picture.
First of all, it leaves out polarization effects, and second
it does not distinguish between a classical object with a cross 
section $\sigma$
(such as a classical oscillating dipole \cite{jackson}) 
and a quantum object 
with the same cross section (such as a $J=0$ to $J=1$ transition in an atom).
Here we take a closer theoretical look at the resonant interaction
of a two-level atom with strongly focused light.

To start with earlier theoretical work,
in Refs.~\cite{carm1,carm2} 
a standard quantum-optical version of input-output 
theory (see \cite{io}, and also \cite{gardiner}) 
was used to calculate the effects of light scattering 
off of a single atom in free space. The description of the light waves
in the model has a 1-dimensional 
character but one may expect  
a full 3-dimensional calculation to be necessary if 
the incoming light is strongly focused. 
Here we show how 
the input-output
theory can be rephrased to include the full 3-dimensional description of the 
input and output light beams and yet keep all characteristics of the 
simpler 1-dimensional model. 

In other work \cite{focus1,focus2} on the same subject
exact 3-dimensional solutions of the 
Maxwell equations were constructed by expanding the field in a complete set of 
functions that are well-suited to describe a cylindrically symmetric beam.
In the present paper we use a different method, and apply the results of
Refs.~\cite{quabis1,quabis2}, where the standard
Debye approximation is used to construct exact expressions for strongly 
focused light beams. The authors
find that a radially polarized
beam of light produces a field
that is longitudinally polarized in the focal spot and that
is focused down to a very small 
spot size, namely $A\approx 0.1\lambda^2$. It is worth noting
that this area is smaller than $\sigma$ by almost a factor of 5.

One may wonder what the strongest focusing possible is. It turns
out that that 
depends on one's definition. The smallest spot size is one criterion, 
but another,
which is the most relevant for our discussion, is to find
the maximum possible electric field intensity in the origin
given a fixed power for an incoming beam.
That maximum is known to be achieved by an electric dipole 
wave \cite{sheppard1}. For 
illumination with a finite numerical aperture one still obtains
the maximum 
by an electric dipole wave, but the optimum then depends
on the polarization, see \cite{sheppard2} and also Section V. 
In any case, a measure of how focused a given light beam is, is thus
given
by its overlap with the appropriate dipole wave. 

Furthermore,
for the interaction of light with an atom in the
usual long-wavelength approximation \cite{cohen}, 
one may expand the interaction in a 
multipole series. To lowest order, the atom interacts through
an electric dipole interaction; hence, to lowest order
the only types of waves interacting 
with the atom are, again, electric dipole waves.
Thus, it makes sense for two reasons to expand the field in multipole 
waves around the origin (which is where the atom is assumed to be).
In addition, we note here that a multipole expansion
is also convenient for the calculation of the field 
distribution in the focal region \cite{sheppard3}.

This paper is organized as follows:
In Section II
we review the essential properties of multipole waves, and in Section III
we consider the interaction of an atom with a field expanded in multipole
waves. In Section IV we calculate two characteristic
quantities of the scattered light as functions of the
location of the photodetector: one is the intensity pattern
and one is the photon statistics.
The former quantity is classical in some sense, as a quantum object
never gives rise to an intensity pattern that cannot be also 
obtained from the Maxwell equations and a classical
scatterer. The photon statistics, on the other hand, does provide a 
quantum signature of the scattering process. In Section V
we consider the (longitudinally polarized) 
focused light beams discussed in \cite{quabis1,quabis2}
 and calculate
their overlap with the appropriate dipole wave. We compare the results to
similar known results on transversely polarized waves.
We conclude by
raising several open questions concerning
various aspects of focused light that are not
treated in the main text.

\section{Multipole waves}
When quantizing the electromagnetic (EM) field, we may choose any complete 
set of mode functions to expand the electric and magnetic fields in.
The usual choice is to take plane waves, but, as mentioned above, 
for the description of the 
interaction of radiation with atoms in the dipole approximation
multipole waves \cite{cohen} are a good alternative.

Multipole waves are eigenfunctions of commuting Hermitian operators such 
that the excitations of such modes (i.e., the photons) possess definite 
amounts of energy $E$ $[E=\hbar \omega]$, total angular momentum $\vec{J}^2$ 
$[\vec{J}^2=J(J+1)\hbar^2]$, angular momentum $J_z$ in the $z$ direction 
$[J_z=M\hbar]$, and parity $P$ [either $(-1)^{J+1}$ (written here as $P=X$) 
or $(-1)^J$ (denoted by $P=Z$)]. 
We thus may expand the electric and magnetic field operators as
\begin{eqnarray}
\vec{E}(\vec{r})&=&\int {\rm d} \omega \sum_\nu
{\cal N}_\omega
\vec{\Phi}_{\omega \nu}(\vec{r})a_{\omega \nu} +H.c. \nonumber\\
c\vec{B}(\vec{r})&=&\int {\rm d} \omega \sum_\nu
{\cal N}_\omega
\vec{\Psi}_{\omega \nu}(\vec{r})a_{\omega \nu}  +H.c,
\end{eqnarray}
where we abbreviated the set of discrete quantum numbers 
as $(J,M,P)=:\nu$ and
$\sum_\nu:= \sum_{J=1}^\infty \sum_{M=-J}^J\sum_{P=X}^Z$.
Furthermore $a^\dagger_{\omega\nu}
$ and $a_{\omega\nu}$
are the creation and annihilation operators for the mode with the 
corresponding eigenvalues $\omega$ and $\nu$.
The functions $\vec{\Phi}$ are normalized to
\begin{eqnarray}
\int {\rm d}\vec{r} \vec{\Phi}^*_{\omega \nu}(\vec{r})\cdot
\vec{\Phi}_{\omega'\nu'}(\vec{r})=(2\pi)^3\delta(\omega-\omega')
\delta_{\nu\nu'}
\end{eqnarray}
and the same for $\vec{\Psi}$. Here we used the obvious
abbreviation $\delta_{\nu\nu'}=\delta_{JJ'}\delta_{MM'}\delta_{PP'}$. 
The normalization factor
\begin{equation}
{\cal N}_\omega
=\left[ \frac{\hbar \omega}{2\epsilon_0(2\pi)^3}  \right]^{1/2}
\end{equation}
is chosen such that the free-field Hamiltonian takes the familiar form
\begin{eqnarray}
H&=&\frac{\epsilon_0}{2}\int {\rm d}\vec{r} 
\left[
\vec{E}^2(\vec{r})+c^2\vec{B}^2(\vec{r})\right]
\nonumber\\
&=&\int {\rm d}\omega \hbar\omega   \sum_\nu \big[ a^{\dagger}_{\omega \nu }
a_{\omega \nu }+\frac{1}{2}\big].
\end{eqnarray}
The expressions for the multipole waves are somewhat  complicated (see
\cite{cohen}), but fortunately we will only need  the values in the
origin (which is where the atom is assumed to be), and in the far
field, at a distance $r$ from the atom with $r\omega_A/c\gg 1$ with
$\omega_A$ the relevant atomic resonance frequency.  In the origin,
the only waves for which the electric field (with which the atom interacts
to first approximation)
is nonzero, are the electric dipole waves.  They
are the waves with quantum numbers $J=1$ and even parity
(corresponding to $P=X$). The other quantum numbers $\omega$ are
arbitrary, and $M$ takes on one of the values $M=\pm 1,0$.  In
$\vec{r}=0$ the dipole waves take on the values \cite{cohen}
\begin{equation}
\vec{\Phi}(0)_{\omega 1MX}=
i\frac{\omega}{c^{3/2}}\left[\frac{8\pi}{3}\right]^{1/2}\hat{u}_M,
\end{equation}
where we explicitly displayed the dipole quantum numbers
$\nu=(J=1,M,P=X)$. The unit vectors
\begin{equation}
\hat{u}_0=\hat{z};\,\,\, \hat{u}_{\pm 1}=(-i\hat{y}\mp
\hat{x})/\sqrt{2}
\end{equation}
are the standard unit circular vectors.  In the far field, a dipole
field $\vec{\Phi}_{\omega 1MX}(\vec{r})$ reduces to \cite{cohen}
\begin{equation}
\vec{\Phi}_{\omega 1MX}(\vec{r})\rightarrow
i\frac{\cos(k_0r-\pi/2)}{r}\left[\frac{6\pi}{
c}\right]^{1/2}[\hat{u}_M-(\hat{u}_M\cdot \hat{r})\hat{r}],
\end{equation}
with $k_0=\omega/c$.

For later use it is also convenient to have the Fourier transform of a
dipole wave at our disposal. It factorizes into the product of a
function of the length of $\vec{k}$ and a function of  the unit vector
$\hat{\kappa}=\vec{k}/k$,
\begin{equation}
\vec{\Phi}_{k_0
1MX}(\vec{k})=\frac{1}{k_0}\sqrt{\frac{3}{8\pi}}\delta(k-k_0)
[\hat{u}_M-(\hat{u}_M\cdot \hat{\kappa})\hat{\kappa}].
\end{equation}
Since we will consider (quasi-)monochromatic waves the
 $\hat{\kappa}$-dependent 
part is the most relevant for our purposes:
\begin{equation}\label{Fourier}
\vec{\Phi}_{M}(\hat{\kappa})= \sqrt{\frac{3}{8\pi}}
[\hat{u}_M-(\hat{u}_M\cdot \hat{\kappa})\hat{\kappa}],
\end{equation}
which is normalized to
\begin{equation}
\int {\rm d}^2\hat{\kappa}| \vec{\Phi}_{M}(\hat{\kappa})|^2=1.
\end{equation}
\section{Atom-field interaction}
Assume we have a two-level atom with a fixed dipole moment
$\vec{d}=d\hat{u}_K$, with $\hat{u}_K$ with $K=0,\pm 1$ one of the
unit circular vectors.  Through the standard electric dipole coupling,
$H_{{\rm int}}=-\vec{d}\cdot\vec{E}$, one gets the following
interaction Hamiltonian in the rotating-wave approximation
\begin{equation}\label{Hint}
H_{{\rm int}}=i\hbar \int {\rm d}\omega \kappa(\omega)
[b^{\dagger}_{\omega}\sigma^- -\sigma^+b_{\omega}],
\end{equation}
where
\begin{equation}
\kappa(\omega)= |\vec{d}\cdot\vec{\phi}(0)_{\omega 1KX}|=
\left[\frac{d^2\omega^3}{6\pi^2\hbar\epsilon_0c^3}\right]^{1/2}.
\end{equation}
Here we defined new creation and annihilation operators
$b^{\dagger}_\omega$ and $b_\omega$ describing the electric dipole
wave that is coupled to the atom, with quantum numbers $J=1,M=K,P=X$
and arbitrary $\omega$.  We denote the discrete quantum numbers by
$\nu=\nu_0$.  Eq.~(\ref{Hint}) is the same interaction Hamiltonian
as that studied in Refs.\cite{carm1,carm2}, but its validity goes beyond a
1-dimensional model by virtue of having expanded the EM field in
multipole waves, as was already mentioned before in
Ref.~\cite{gardiner}.

We can now write down the Heisenberg equations of motion for atomic
and field operators. The equation for $b_{\omega}(t)$ is easily solved
to give
\begin{equation}\label{b}
b_{\omega}(t)=e^{-i\omega(t-t_0)}b_0(\omega)+\kappa(\omega)\int {\rm
d}t' e^{-i\omega(t-t')}\sigma^-(t'),
\end{equation}
where $t_0$ is a time in the far past ($t_0<t$) and
$b_0(\omega)=b_\omega(t_0)$.  The equations for the mode operators for
all remaining modes $(\omega,\nu)$ contain just the free evolution, so
that
\begin{equation}\label{a}
a_{\omega,\nu}(t)=e^{-i\omega(t-t_0)}a_0(\omega,\nu).
\end{equation}
The standard Markov approximation now consists of assuming that
$\kappa(\omega)$ is more or less constant in the relevant frequence
range around the atomic frequency $\omega_A$, and we approximate
\begin{equation}
\kappa(\omega)\approx \left[
\frac{d^2\omega_A^3}{6\pi^2\hbar\epsilon_0c^3}\right]^{1/2} =:
\left[\frac{\Gamma}{2\pi}\right]^{1/2},
\end{equation}
with $\Gamma$ the spontaneous emission rate constant. (Thus, one finds
the same expression for $\Gamma$ as in a plane-wave expansion.)
Substituting this into the equations for the atomic operators gives
\begin{eqnarray}\label{Heis}
\dot{\sigma}^-&=&-i\omega_A\sigma^- -\frac{\Gamma}{2}\sigma^-+
\sqrt{\Gamma}\sigma_zb_{{\rm in}}(t) \nonumber\\
\dot{\sigma}_z&=&-\Gamma(1+\sigma_z)-2\sqrt{\Gamma}\sigma^+b_{{\rm
in}}(t) -2\sqrt{\Gamma}\sigma^-b^{\dagger}_{{\rm in}}(t).\nonumber\\
\end{eqnarray}
(The last equation also follows from $\sigma_z=2\sigma^+\sigma^--1$.)
Here we defined the input (justifying its name from the fact that this
field drives the atom) operator \cite{io}
\begin{equation}
b_{{\rm in}}(t)=\frac{1}{\sqrt{2\pi}}\int {\rm d}\omega
e^{-i\omega(t-t_0)} b_0(\omega),
\end{equation}
which satisfies the commutation relation
\begin{equation}
[b_{{\rm in}}(t),b^{\dagger}_{{\rm in}}(t')]=\delta(t-t').
\end{equation}
Substituting Eqs.~(\ref{b}) and (\ref{a}) into the expansion of the
electric field yields the usual separation into source and free parts,
$\vec{E}=\vec{E}_{{\rm free}}+\vec{E}_{{\rm source}}$. In the far field
\begin{equation}
\vec{E}^{(+)}_{{\rm source}}(\vec{r},t)\rightarrow
\frac{d\omega_A^2}{4\pi\epsilon_0c^2} \frac{ \hat{u}_K-(\hat{u}_K\cdot
\hat{r})\hat{r} }{r} \sigma^-(t-r/c),
\end{equation}
where for convenience we will only display the positive-frequency
parts of the fields from now on; 
the negative-frequency part is just the Hermitian
conjugate of the positive-frequency part,
$\vec{E}^{(-)}=(\vec{E}^{(+)})^{\dagger}$.  Causality is obeyed as the
atomic operator must be evaluated at a retarded time $t-r/c$ with
$r=|\vec{r}|$ the distance from the atom.

The free field contains two terms, one corresponding to the relevant
dipole mode, the other to all remaining modes. The former depends only
on the operator $b_0(\omega)$, the latter on $a_0(\omega,\nu)$,
according to (again valid in the far field)
\begin{eqnarray}
\vec{E}^{(+)}_{{\rm free}}(\vec{r},t)&\rightarrow& \int {\rm d}\omega
b_0(\omega) e^{-i\omega(t-t_0)} \cos(kr-\pi/2) \nonumber\\ &&\times
\left[ \frac{\hbar \omega}{2\epsilon_0(2\pi)^3}
\right]^{1/2}\big(\frac{6\pi}{c}\big)^{1/2}i\frac{
\hat{u}_K-(\hat{u}_K\cdot \hat{r})\hat{r}}{r}\nonumber\\
&+&\sum_{\nu\neq \nu_0}\int {\rm d}\omega a_0(\omega,\nu)
e^{-i\omega(t-t_0)}\cos(kr-J\pi/2)  \nonumber\\ &&\times \left[
\frac{\hbar \omega}{2\epsilon_0(2\pi)^3}
\right]^{1/2}\big(\frac{6\pi}{c}\big)^{1/2} (i)^J
\frac{\vec{P}_{JM}(\hat{r}) }{r},
\end{eqnarray}
where $\vec{P}_{JM}(\hat{r})$ is a transverse vector (perpendicular to
$\vec{r}$), whose form depends only on the quantum numbers $J$ and
$M$, and which is a function of the unit vector $\hat{r}$
\cite{cohen}.  Again making use of a Markov approximation, we can
rewrite this in terms of the input fields, evaluated at the earlier
time $t-r/c$,
\begin{eqnarray}
\vec{E}^{(+)}_{{\rm free}}(\vec{r},t)&\rightarrow&  \mu b_{{\rm
in}}(t-r/c) \frac{ \hat{u}_K-(\hat{u}_K\cdot
\hat{r})\hat{r}}{r}\nonumber\\ &+&\sum_{\nu\neq\nu_0}\mu a_{{\rm
in}}^{\nu}(t-r/c)  \frac{\vec{P}_{JM}(\hat{r}) }{r},
\end{eqnarray}
where
\begin{equation}
\mu=\big[\frac{3\hbar \omega_A}{16\pi \epsilon_0 c}\big]^{1/2}.
\end{equation}
We introduced here another set of input fields (although here these
fields do not drive the atom in any way),
\begin{equation}
a_{{\rm in}}^{\nu}(t)=\int {\rm d}\omega a_0(\omega,\nu)
e^{-i\omega(t-t_0)},
\end{equation}
with commutation relations
\begin{equation}
[a^\nu_{{\rm in}}(t),a^{\nu\dagger}_{{\rm in}}(t')]=\delta(t-t').
\end{equation}
The total field becomes then
\begin{eqnarray}
\vec{E}^{(+)}(\vec{r},t)&\rightarrow&
\big[\sqrt{\Gamma}\sigma^-(t-r/c) + b_{{\rm in}}(t-r/c)\big]
\nonumber\\ &&\times\mu \frac{ \hat{u}_K-(\hat{u}_K\cdot
\hat{r})\hat{r}}{r} \nonumber\\ &+& \sum_{\nu\neq\nu_0}  a_{{\rm
in}}^{\nu}(t-r/c)\mu \frac{\vec{P}_{JM}(\hat{r})}{r}
\end{eqnarray}
In the first line one recognizes the standard expression
\cite[Eq. (2.22)]{io} for the output field operator
\begin{equation}
b_{{\rm out}}(t)=b_{{\rm in}}(t)+\sqrt{\Gamma}\sigma^-(t).
\end{equation}
This operator, too, satisfies
\begin{equation}
[b_{{\rm out}}(t),b^{\dagger}_{{\rm out}}(t')]=\delta(t-t').
\end{equation}

\section{Photon flux and statistics}
In the following we suppose we measure the flux of the output field
and its statistics for a fixed polarization $\hat{\epsilon}$  at some
fixed position $\vec{r}=\vec{R}$ in the far field.  
This is not an essential assumption, and
we could easily define quantities
similar to the ones defined below
for every polarization component detected.

It is convenient
to introduce yet another input field operator
\begin{equation}
a_{{\rm in}}(t)=\frac{1}{{\cal P}}\sum_{\nu\neq\nu_0}a^\nu_{{\rm
in}}(t) \vec{P}_{JM}(\hat{R})\cdot\hat{\epsilon},
\end{equation}
where the presence of the geometric factor
\begin{equation}
{\cal
P}^2=\sum_{\nu\neq\nu_0}|\vec{P}_{JM}(\hat{R})\cdot\hat{\epsilon}|^2
\end{equation}
ensures that the relation
\begin{equation}
[a_{{\rm in}}(t),a^{\dagger}_{{\rm in}}(t')]=\delta(t-t')
\end{equation} 
holds.  Defining a similar geometric factor for the dipole field $b$,
\begin{equation}
{\cal
D}=\hat{u}_K\cdot\hat{\epsilon}-(\hat{u}_K\cdot\hat{R})
(\hat{R}\cdot\hat{\epsilon}),
\end{equation}
the detection operator 
can then be compactly written in terms of the operator
\begin{equation}
C(t)={\cal P}a_{{\rm in}}(t)+{\cal D}[b_{{\rm
in}}(t))+\sqrt{\Gamma} \sigma^-(t)],
\end{equation}
with $R=|\vec{R}|$: namely, we get
\begin{equation}
\hat{\epsilon}\cdot\vec{E}^{(+)}(\vec{R},t) \rightarrow \frac{\mu}{R}
C(t-R/c).
\end{equation}
We should note here that the operators appearing in $C$ do not all
commute. In particular, $b_{{\rm in}}(t)$ does not commute with the
atomic operators, but $a_{{\rm in}}(t)$ does.

We can define two quantities of interest: a photon flux operator (with
the dimension of a rate)
\begin{equation}
F=\langle C^{\dagger}(t)C(t)\rangle,
\end{equation}
and the second-order intensity correlation function at time zero,
$g^{(2)}(0)$,
\begin{eqnarray}
g^{(2)}(0)&=&\frac{G^{(2)}(0)} {F^2}\nonumber\\ G^{(2)}(0)&=& \langle
C^{\dagger 2}(t) C^2(t)\rangle.
\end{eqnarray}
The ordering of the noncommuting operators $b_{{\rm in}}$ and the
atomic operators matters here. We can make use of a theorem given in
\cite{vogel} that states that in expressions such as those for $F$ and
$g^{(2)}(0)$ we can place  $\vec{E}^{(+)}_{{\rm source}}$  to the left of
$\vec{E}^{(+)}_{{\rm free}}$ and $\vec{E}^{(-)}_{{\rm source}}$  to
the right of $\vec{E}^{(-)}_{{\rm free}}$. That is, we can place all
operators $b_{{\rm in}}$ to the right of atomic operators and
$b^{\dagger}_{{\rm in}}$ to the left.

This way, we can easily calculate these quantities in special cases of
interest.  In all cases we assume the field illuminating the atom has
a central frequency $\omega_L$ with a bandwidth $B$ sufficiently
narrow so that $B$ is smaller than other rates in the problem, $B\ll
\Gamma$ and $B \ll c/R$.  Often, we will be interested in the
steady-state (in a frame rotating at the laser frequency $\omega_L$)
solution.  Taking expectation values and moving to a frame rotating at
$\omega_L$, transforms  the Heisenberg equations (\ref{Heis}) into the
optical Bloch equations
\begin{eqnarray}\label{Bloch}
\langle\dot{\sigma}^-\rangle&=&\big(i\Delta-\frac{\Gamma}{2}\big)
\langle\sigma^-\rangle +\sqrt{\Gamma}\langle\sigma_zb_{{\rm
in}}(t)\rangle \nonumber\\
\langle\dot{\sigma}_z\rangle&=&-\Gamma(1+\langle\sigma_z\rangle)-
2\sqrt{\Gamma}\langle\sigma^+b_{{\rm in}}(t)\rangle
-2\sqrt{\Gamma}\langle b^{\dagger}_{{\rm
in}}(t)\sigma^-\rangle,\nonumber\\
\end{eqnarray}
where the laser detuning from resonance is $\Delta=\omega_L-\omega_A$.
For later use we note that in the steady state the second equation
gives
\begin{equation}\label{steady}
\sqrt{\Gamma}\big[\langle b^{\dagger}_{{\rm in}} \sigma^- \rangle
+\langle \sigma^+ b_{{\rm in}} \rangle\big] +\Gamma\langle
\sigma^+\sigma^- \rangle=0
\end{equation}
\subsection{Coherent states}
We are interested in the case of illumination with a field in a
coherent state.   One reason is that this corresponds to a laser
field, another is that the statistics of the incoming light is then
Poissonian, so that possible non-classical statistics arise from the
scattering process, not from the incoming light.

Hence, we assume the input fields satisfy
\begin{eqnarray}
b_{{\rm in}}(t)|\beta\rangle&=&\beta \exp(-i\omega_L t) |\beta\rangle,
\nonumber\\ a_{{\rm in}}(t)|\alpha\rangle&=&\alpha \exp(-i\omega_L t)
|\alpha\rangle.
\end{eqnarray}
Thus, the input field is described by just two complex amplitudes, one
for the relevant dipole part, one for the rest.  We also define
$\eta$, a dimensionless number  relating the total  amplitude of the
free field at the detection point  to the contribution of the dipole
field, as
\begin{equation}
{\cal D}\eta\beta={\cal P}\alpha+{\cal D}\beta.
\end{equation}
Below we will connect these quantities to the overlaps with dipole and
other multipole waves.  The optical Bloch equations (\ref{Bloch})
depend only on the amplitude  $\beta$
\begin{eqnarray}\label{Blochc}
\langle\dot{\sigma}^-\rangle&=&\big(i\Delta-\frac{\Gamma}{2}\big)
\langle\sigma^-\rangle +\sqrt{\Gamma}\beta\langle\sigma_z\rangle
\nonumber\\
\langle\dot{\sigma}_z\rangle&=&-\Gamma(1+\langle\sigma_z\rangle)-
2\sqrt{\Gamma}\beta\langle\sigma^+\rangle
-2\sqrt{\Gamma}\beta^*\langle\sigma^-\rangle.\nonumber\\
\end{eqnarray}
The steady-state solution is
\begin{eqnarray}
\sqrt{\Gamma}\langle \sigma^-\rangle_s&=&
\frac{-2\beta(1+i\delta)}{1+\delta^2+8|\beta|^2/\Gamma},\nonumber\\
\langle \sigma_z\rangle_s&=&
\frac{-(1+\delta^2)}{1+\delta^2+8|\beta|^2/\Gamma},
\end{eqnarray}
where we defined the dimensionless detuning $\delta=2\Delta/\Gamma$.
It is now straightforward to calculate the quantities $F$ and
$g^{(2)}(0)$.

First, we compare our result with that of Ref.~\cite{carm1,carm2} for
resonant excitation ($\delta=0$). We find for the flux $F$
\begin{equation}
F\propto \frac{(1-2/|\eta|)^2 +8|\beta|^2/\Gamma}{1+8|\beta|^2/\Gamma}.
\end{equation}
Compare this with the expression from \cite{carm2}, Eq.~(30),
\begin{equation}
F={\cal R}\frac{\big(1-2\gamma_S/\gamma\big)^2 +8{\cal
R}\gamma_S/\gamma^2}{1+8{\cal R}\gamma_S/\gamma^2},
\end{equation}
with ${\cal R}$ the total incident flux, $\gamma_S$ the spontaneous
emission rate into the solid angle subtended by the incident beam and
$\gamma$ the total spontaneous emission rate (corresponding to our
$\Gamma$). Extreme focusing corresponds to $\gamma_S=\gamma/2$.  We
can then make the identifications
\begin{eqnarray}\label{compare}
{\cal R}&\leftrightarrow&|\eta||\beta|^2,\nonumber\\
\gamma_S/\gamma&\leftrightarrow&1/|\eta|.
\end{eqnarray}
This comparison, though, is not perfect. In our case $\eta$ is complex
and can, in principle, take on any value, but $0\leq
2\gamma_S/\gamma\leq 2$ (and with light coming from 1 direction only,
one even has $2\gamma_S/\gamma\leq 1$).

The strongest effects on photon statistics occur in the weak driving
limit $|\beta|^2\ll \Gamma$, ---in the strong driving limit the atom
saturates and the output field will display Poissonian statistics---
and on resonance.  We therefore consider the special case of weak
on-resonance excitation, and obtain
\begin{eqnarray}
F&=&|{\cal D}|^2|\beta|^2|\eta-2|^2,\nonumber\\
g^{(2)}(0)&=&\frac{|\eta|^2|\eta-4|^2}{|\eta-2|^4},
\end{eqnarray}
This result of $g^{(2)}(0)$ is plotted in Figure~1. Remarkably, the
plot  is very similar to that obtained in Ref.~\cite{focus2} for
illumination with Gaussian beams and detection in the forward
direction (see Fig.~8 there).  In the latter case, the result is
plotted as a function of the beam waist.  The remarkable aspect is,
though, that the Gaussian beams are in fact no  longer solutions to
the Maxwell equations as the focusing conditions are too strong for
the paraxial approximation to be valid.  Here, on the other hand, the
result is not an approximation, but is plotted as a function of
$|\eta|$.
\begin{figure}
\includegraphics[scale=0.4]{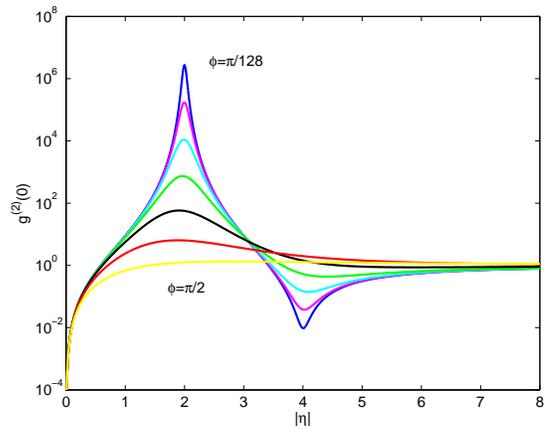}
\caption{$g^{(2)}(0)$, in the weak-driving limit and on resonance, as
a function of $|\eta|$ for various values of the  complex phase $\phi$
of $\eta$: $\phi=\pi/(2n)$ for $n=1\ldots 7$. }
\end{figure}

For reference we also give the flux of the output dipole field at
position $\vec{R}$ in the {\em absence} of the atom: $F_0=|{\cal
D}\beta|^2$.  This leads us to the following observations in special
cases
\begin{itemize}

\item[$\eta=1$] If one illuminates the atom with only the dipole
field, so that $\eta=1$, one has the strongest possible focusing.  All
light is coupled to the atom, and we note that the correspondence
relation (\ref{compare}) shows that this case corresponds to
$2\gamma_S/\gamma=2$ in terms of Carmichael's parameters, indicating
illumination with light that fills the full $4\pi$ solid angle.
Moreover, this case is the only one where $\eta$ and thereby the flux
and photon statistics do not depend on the detection point.

For $\eta=1$ the output flux equals the input flux: $\langle
b^{\dagger}_{{\rm out}}b_{{\rm out}}\rangle =\langle b^{\dagger}_{{\rm
in}}b_{{\rm in}}\rangle$, and $F=F_0$.  One way to understand this is
to note that while $\langle b_{{\rm in}}\rangle=\beta$, we have for
the source field $\sqrt{\Gamma}\langle\sigma^-\rangle=-2\beta$. Thus
the expectation value of the total field is $-\beta$. However, this
does not mean that a $\pi$ phase change is the only difference.

In fact, the statistics of the light has been affected by the presence
of the atom. The output photons are bunched as $g^{(2)}(0)=9$, even
though the input field dispayed Poissonian statistics. The explanation
for this effect is similar to that given in \cite{carm1,carm2} for
strong bunching: the atom cannot absorb any photons if it is in the
excited state, but can if it is in the ground state. The detection of
a photon thus makes it more likely the atom would be found in the
excited state, which in turn makes it more likely to detect a second
photon, namely one emitted by the atom.

\item[$\eta=0$] When the {\em free} fields interfere destructively at
the observation point (so that $\eta=0$), the field there arises
solely from the atom's fluorescence. That light, as is well-known, is
anti-bunched and $g^{(2)}(0)=0$ (as we are in the weak driving
limit). (The reason is simple, the detection operator is $C={\cal
D}\sqrt{\Gamma}\sigma^-$ and applying this operator twice yields
zero.)  The flux $F$ is $F=4F_0$ as a result of the source field being
twice as strong as the input dipole field,
$\sqrt{\Gamma}\langle\sigma^-\rangle=-2\beta$.

\item[$\eta\rightarrow 2$] When $\eta=2$ one finds the largest
bunching effect, with $g^{(2)}(0)\rightarrow\infty$. This occurs simply
because the total intensity there vanishes, $F\rightarrow 0$, as the
source field (of amplitude $-2\beta$)  destructively interferes with
the free field (with dipole wave  and the rest each contributing an
amplitude $\beta$).

\item[$\eta=4$] When $\eta=4$  the total detected flux is equal to
$F=4F_0$  with now the free fields contributing $4\beta$ to the
amplitude, the source field subtracting $2\beta$, as before.
Interestingly, we again have completely anti-bunched light,
$g^{(2)}(0)=0$ (at least in the low-intensity limit).  Here is why (in a
quantum-trajectory picture)  the light is anti-bunched in this case:
The detection operator is effectively $C={\cal D}
(4\beta+\sqrt{\Gamma}\sigma^-)$, since the state of the free radiation
field (in the detection point) is a coherent state with amplitude
$4\beta$.  The steady state of the atom in between photodetection
events is $|\psi\rangle=|g\rangle-2\beta/\sqrt{\Gamma}|e\rangle$
(valid to first order order in the small parameter
$\beta/\sqrt{\Gamma}$).  After the first detection of a photon, we
collapse the state onto $C|\psi\rangle\propto
|g\rangle-4\beta/\sqrt{\Gamma}|e\rangle=:|\phi\rangle$.  The
probability rate of another photon detection  is proportional to the
norm of the wave function $C|\phi\rangle=
-16\beta^2/\sqrt{\Gamma}|e\rangle$, which is only of order $\Gamma
(|\beta|^2/\Gamma)^2$.  In words, the two paths to produce a photon
after the first photodetection event (one from the laser field, the
other from the atom) interfere almost completely destructively.
Hence, $g^{(2)}(0)\rightarrow 0$ in the weak driving limit.

We note here that this type of anti-bunching is connected to a
collapse of the atom to the {\em excited} state after the first
photodetection,  in contrast to the case of pure fluorescence, where
the atom is collapsed into the ground state.

\item[$\eta\rightarrow \infty$] If the field not coupled to the atom
is large, the photon statistics of the total field will be dominated
by that field, leading to Poissonian light with $g^{(2)}(0)\approx 1$.

\end{itemize}

\section{The Debye approximation and dipole waves}
The overlap of the incoming field with the appropriate dipole wave is
clearly the most crucial quantity, as only that part interacts with
the atom.  For instance, if we denote that overlap by ${\cal O}_d$,
then it easy to see that the following two relations, involving
parameters used before,  hold:
\begin{eqnarray}
{\cal O}_d&=&\frac{\beta}{\sqrt{|\alpha|^2+|\beta|^2}}\nonumber\\
|\eta-1|^2&=&\frac{{\cal P}^2}{{\cal D}^2} \frac{1-|{\cal
O}_d|^2}{|{\cal O}_d|^2}.
\end{eqnarray}
One question relevant in practice is, what is the largest overlap
possible for a given numerical aperture?  In order to answer this
question we revert to the Debye approximation.  In that approximation
the field in a focus, resulting from a high numerical aperture, is
expanded in plane waves (hence, the resulting expressions are
conveniently given as Fourier transforms). The approximation consists
in taking into account only the geometric optics rays from the
aperture, thus leaving out edge effects. For a recent discussion
of high-aperture beams and more background information, 
see \cite{sheppard4}.

In Refs~\cite{quabis1,quabis2} this approximation is used to calculate
the intensity profile of  a particularly strongly focused type of
waves that were also generated in an actual experiment.  Those waves
are {\em longitudinally} polarized in the focal spot.  As before, in
reciprocal space the wave function factorize into a  delta function
$\delta(k-k_0)$ and a part depending only on the unit vector
$\hat{\kappa}$. We only need the latter part, which we denote by
$\vec{\chi}$.  In spherical coordinates $(\alpha,\beta)$ for
$\hat{\kappa}$ one gets
\begin{eqnarray}
\vec{\chi}(\alpha,\beta)&=& \frac{1}{\sqrt{{\cal
N}}}A(\alpha)\hat{p}(\alpha,\beta)
\,\, {\rm for}\,\,  \alpha\leq \theta\nonumber\\
&=&0\,\, {\rm otherwise},
\end{eqnarray}
with $\hat{p}(\alpha,\beta)$ a 
unit vector indicating the direction of the electric
field,
\begin{eqnarray}
\hat{p}(\alpha,\beta)= \left(
\begin{array}{c}
\cos\alpha\cos\beta\\ \cos\alpha\sin\beta\\ \sin\alpha.
\end{array}
\right)
\end{eqnarray}
The previous expressions all follow directly from Eq.~(9) in
Ref.~\cite{quabis2}.  Here $\theta$ is related to the numerical
aperture by $NA=\sin\theta$, and the normalization factor ${\cal N}$
is given by
\begin{equation}
{\cal N}=2\pi \int_0^\theta {\rm d}\alpha \sin\alpha |A(\alpha)|^2.
\end{equation}
The factor $A(\alpha)$ is determined by the input field on the
lens. For the  class of light beams studied in \cite{quabis1,quabis2},
\begin{equation}\label{quab}
A(\alpha)=\sin\alpha \sqrt{|\cos\alpha|}\exp(-a^2\sin^2\alpha),
\end{equation}
with $a=f/w_0$ the ratio between the focal length $f$ of the lens and
the waist $w_0$ of the incoming Gaussian beam. The factor 
$\sqrt{|\cos\alpha|}$ is typical for an aplanatic lens system.

In order to determine the overlap we need the Fourier transform
of the relevant dipole wave. Here we need $K=0$ and $\vec{\Phi}_0$, 
since the
polarization in the origin 
is directed along $\hat{u}_0=\hat{z}$.  From
Eq.~(\ref{Fourier}) we read off
\begin{eqnarray}
\vec{\Phi}_0(\alpha,\beta)=\sqrt{\frac{3}{8\pi}}\sin\alpha
\hat{p}(\alpha,\beta)
\end{eqnarray}
normalized to
\begin{equation}
\int_0^{2\pi}{\rm d}\beta \int_0^{\pi}{\rm d}\alpha\sin\alpha
|\vec{\Phi}_0(\alpha,\beta)|^2=1.
\end{equation}
The overlap between $\vec{\chi}$ and $\vec{\Phi}_0$ is then
\begin{equation}
{\cal O}_d= 2\pi \int_0^\theta  {\rm d}\alpha \sin\alpha
\vec{\chi}\cdot\vec{\Phi}_0 =2\pi \int_0^\theta  {\rm d}\alpha
\sin^2\alpha A(\alpha)/\sqrt{{\cal N}}.
\end{equation}
This overlap has been calculated numerically as a function of $\theta$, and
the results are plotted in Fig.~2.
\begin{figure}
\includegraphics[scale=0.4]{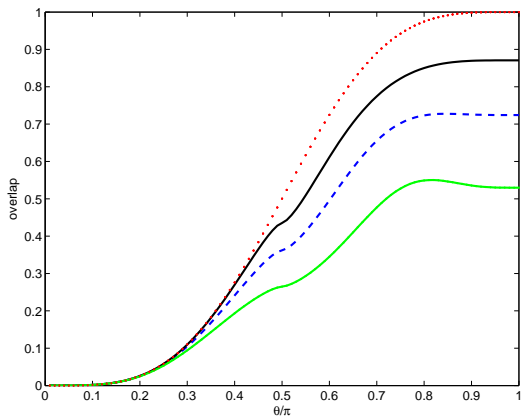}
\caption{Dipole wave content $p=|{\cal O}_d|^2$, 
for the longitudinally polarized waves of Eq.~(\protect{\ref{quab}})
for several values of
$a=f/w_0$: the three bottom
curves correspond to $a=2,1,0$, respectively. The top curve
gives the maximum possible overlap $|{\cal O}_{{\rm max}}|^2$.}
\end{figure}
The plot also shows the maximum possible overlap for given numerical
aperture.  That maximum is achieved when $A(\alpha)\propto
\sin\alpha$. This is most easily seen by defining a scalar product
\begin{equation}
\langle\vec{\Phi},\vec{\Psi}\rangle:= \int_0^\theta
{\rm d}\alpha
\sin\alpha \vec{\Phi}\cdot\vec{\Psi}^*,
\end{equation}
and noting one has to optimize this scalar product  over all
$\vec{\Phi}$ for fixed $\vec{\Psi}=\sin\alpha \hat{p}$.  The maximum
possible overlap with a dipole wave is then
\begin{equation}
|{\cal O}_{{\rm max}}|^2=\frac{1}{2}+
\frac{1}{4}\cos^3\theta-\frac{3}{4}\cos\theta.
\end{equation}
In particular, for $\theta=\pi/2$, which corresponds to the strongest
possible focusing given that the incoming light comes from one
direction, one gets $|{\cal O}_{{\rm max}}|^2=1/2$, while for the
types of beams considered here the best one can do is to go to the
limit $a\rightarrow 0$ when $|{\cal O}_d|^2=64/147\approx 43.5\%$.

The figure also shows that for light coming from one direction
the light beams from Refs.~\cite{quabis1,quabis2}
are very close to the optimum (for fixed polarization) 
in the limit of small $a$, 
but for $\theta>\pi/2$ the distance from the optimum
suddenly increases.

In most optical experiments, however,
light beams in the focus are {\em transversely} polarized.
The overlap of such beams
with a transversely polarized dipole wave (in our notation,
such a linearly
polarized dipole wave would, of course, be a superposition of $\vec{\Phi}_{1}$
and $\vec{\Phi}_{-1}$), has been studied before in 
Refs.~\cite{sheppard1,sheppard2}. It turns out 
(see Fig.~1 in \cite{sheppard1} and Fig.~3 below) 
that for opening angles $\theta$ 
less than $\pi/2$
the overlap with the appropriate dipole waves tends to be larger for 
transverse than for longitudinal polarizations. 
This is a consequence of the latter radiation
pattern having larger side lobes.
This observation is true both for a truncated
dipole wave and for more realistic
light beams. For
$\theta=\pi/2$ the dipole content of truncated dipole waves
is exactly $50\%$ for both polarizations,
and for values of $\theta$ larger than $\pi/2$ 
longitudinally polarized waves have larger overlaps.
This is plotted in Fig.~3.
\begin{figure}
\includegraphics[scale=0.4]{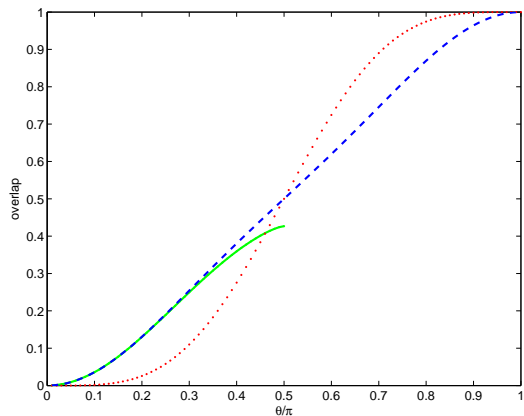}
\caption{Dipole wave content $p=|{\cal O}_d|^2$ 
for various waves.
The dashed curve gives the maximum possible overlap with a transversely
polarized electric dipole wave, obtained for a truncated
dipole wave. The dotted curve gives the same but then 
for longitudinal polarization (and is the same dotted curve
as in Fig.~2). The solid curve corresponds to uniform illumination
with a transverse wave (called ``Sine'' wave in \cite{sheppard2}:
the result given there is valid for $\theta\leq \pi/2$). 
These curves are all obtained from 
Ref.~\cite{sheppard2}.} 
\end{figure}
Let us finally compare the numbers for
more realistic beams:
The transversely polarized wave that results from (nearly) uniform 
illumination
of an aplanatic lens (called ``Sine'' wave in \cite{sheppard2})
has a dipole wave content 
$|{\cal O}_d|^2=32/75\approx 42.7\%$ at $\theta=\pi/2$,
which is in fact slightly smaller than the number quoted above
for the beams described by Eq.~(\ref{quab}) in the limit $a\rightarrow 0$.
\section{Discussion}
We have shown how to reformulate the quasi 1-dimensional theory of
\cite{carm1,carm2} to study photon statistics effects in resonant
light scattering off  an atom in the full 3-dimensional setting, by expanding
the fields in multipole waves \cite{cohen}.  We applied this formalism
to the case of illuminating an atom with strongly focused light, and
found  that there is a single parameter, $\eta$, that contains all the
important information. This is true when one assumes one detects a
single polarization component. Otherwise, for each polarization
component that is detected there is a parameter like $\eta$.

The overlap of the incoming light beam with an electric dipole wave
determines the strength of the atom-light interaction. One expects
that smaller focal spot sizes tend to
correspond to  larger overlaps, as
indeed the dipole wave is the only wave with a nonzero intensity in
the origin. We found a confirmation of this suspicion  in Figure~2,
where the overlaps of a particular class of focused light beams with
very small spot sizes are plotted.  These beams turn out to have almost
the maximum possible overlap, given a fixed numerical aperture (this
is  true within the Debye approximation), and given a {\em longitudinal}
polarization. With this measure,
the more standard case of transverse polarization
leads in fact to even better focusing for opening angles 
$\theta$ less than $\pi/2$,
although the spot sizes is actually larger \cite{quabis1,quabis2}. 
For $\theta>\pi/2$, on the other hand,
longitudinal polarization becomes better than transverse
polarization.

Finally, we note that focused light contains a great deal more
structure than just a small focal spot size. In particular there are
several different types of topological properties that are both robust
(i.e., they do not disappear when boundary conditions are slightly
changed) and generic (i.e., they occur under general, non-special
circumstances).  For a nice discussion see \cite{nye}.  Some of those
properties, such as phase and polarization singularities, occur on
length scales at or even below the  wavelength of the
light. Similarly, around a zero of the intensity  the spectral density
of the light can be singular as well \cite{wolf}.  An open question is
what the quantum signatures are of such topological structures, and
whether one can probe those with an atom or a quantum dot?  An atom is
most likely too small to notice changes of polarization or  spectrum,
even if the change takes place within a wavelength, but a quantum dot
may just be sufficiently large.  In that case one may wonder, 
is there a difference
between a quantum dot seeing red-shifted  and blue-shifted  light
everywhere or seeing red-shifted light in one location and
blue-shifted light in another?  Or similarly, is there a difference
between seeing unpolarized light or seeing horizontal polarization in
one location and vertical polarization in another?
\section*{Acknowledgments}
There is a whole list of people to thank for their valuable input,
comments and discussions:  first H.J.~Kimble for many discussions on
the motivation and ideas for this work, then G.~Leuchs, S.~Quabis, and
R.~Dorn for suggesting that  ``their'' light beams may have a large
overlap with dipole waves (and indeed, they do), for other discussions
on focused light beams, and for their hospitality in Erlangen, and
finally C.J.R.~Sheppard for useful comments 
and for pointing out that the strongest focusing
possible is achieved by (truncated) dipole waves.

\end{document}